\documentclass[
reprint,
amsmath,amssymb,
aps,prd]{revtex4-2}
\usepackage{placeins}
\usepackage{graphicx}
\usepackage{dcolumn}
\usepackage{bm}
\usepackage{hyperref}
\usepackage{xcolor}

\begin{document}

\title{ Deep Learning based discovery of Integrable Systems}

\author{Shailesh Lal}
\email{shaileshlal@bimsa.cn}
\affiliation{Beijing Institute of Mathematical Sciences and Applications, Beijing 101408, China}
\author{Suvajit Majumder}%
\email{majumder.suvajit95@gmail.com}
\affiliation{GoBubble AI Tech Limited, Dubai, UAE\,,}
\affiliation{Department of Mathematics, City University of London, EC1V 0HB, UK}

\author{Evgeny Sobko}
\email{evgenysobko@gmail}
\affiliation{London Institute for Mathematical Sciences, Royal Institution, London, W1S 4BS, UK
}%

\date{\today}

\begin{abstract}

We introduce a novel machine learning  based framework for discovering integrable models. Our approach first employs a synchronized ensemble of neural networks to find high-precision numerical solution to the Yang-Baxter equation within a specified class. Then, using an auxiliary system of algebraic equations, \([Q_2, Q_3] = 0\), and the numerical value of the Hamiltonian obtained via deep learning as a seed, we reconstruct the entire Hamiltonian family, forming an algebraic variety. We illustrate our presentation with three- and four-dimensional spin chains of difference form. Remarkably, all discovered Hamiltonian families form rational varieties.
\end{abstract}
\maketitle
Symmetry is one of the most fundamental and beautiful tools for understanding physics. It guides the development of new theories, enhances physical intuition, and reveals deep underlying mathematical structures. Integrable models are a special class of systems with an infinite group of hidden symmetries, allowing them to be solved exactly at any interaction strength. These models appear in numerous areas of theoretical physics, ranging from quantum mechanical systems like spin chains and lattice models to quantum field theories (QFT) and string theories. They often capture essential non-perturbative phenomena and provide a powerful framework for theoretical exploration. However, discovering new integrable models remains a formidable challenge, with no systematic approach currently available. At the core of integrability lies the Yang-Baxter Equation (YBE), a fundamental constraint that the S-matrix in QFTs and the R-matrix in spin chains and lattice models must satisfy. Solving the YBE is challenging due to its nonlinearity and the enormous number of possible cases to analyze. Over the years, this problem has been approached using various methods. For one, there are algebraic approaches reliant on the symmetries of the R-matrix or Hamiltonian \cite{Kulish:1981gi,Jimbo:1985ua,Bazhanov:1984gu,Bazhanov:1986mu,10.1143/PTP.68.508,Perk:1981nb}. Secondly, methods based on directly solving YBE or differential equations associated with them \cite{vieira2018solving}. The 
third alternative which has recently been developed utilizes the boost operator to generate higher charges and 
impose their commutativity, a constraint also known as the Reshetikhin condition \cite{de2019classifying,deLeeuw:2020ahe,deLeeuw:2020xrw,Corcoran:2023zax}. The last method is quite general in principle; however, in practice, its application is limited to systems with a relatively small number of vertices. Otherwise, the YBE and the algebraic system representing the Reshetikhin condition becomes excessively complex. In order to overcome this significant computational barrier, we had previously introduced a neural network solver, the \texttt{R-Matrix Net} \cite{Lal:2023dkj}. We had further provided experiments which indicated the utility of this method to explore the landscape of integrable models within a fixed setting where a complete classification of solutions was already known \cite{de2019classifying}. However in order to claim discovery of a new model and perform further analysis, numerical method must eventually find a way towards an analytic solution. Then, one may explicitly verify that the criteria for integrability are identically satisfied, thus resolving questions of numerical precision.

In this paper we introduce a novel AI-based  framework for the systematic discovering integrable systems in the exact analytical form. The construction hinges on two key elements, the first one is the \texttt{R-Matrix Net} whose primary goal is to find a high-precision numerical R-matrix from a specified class. The corresponding numerical Hamiltonian is extracted using Equation \eqref{eq:HfromR}. The second key element is the Reshetikhin condition \eqref{eq:q2q3constraint}. For us, this equation is used to extract the exact analytical form of the integrable Hamiltonian family from the initial numerical one. The R-matrix itself can then be obtained using standard analytical methods such as the small spectral-parameter expansion or the Sutherland equation \cite{de2019classifying,deLeeuw:2020ahe,deLeeuw:2020xrw,Corcoran:2023zax} 
We apply this framework to local spin-chains with site space of dimension $d$, and restrict ourselves to R-matrices of difference form. We showcase examples in dimensions $d=3,4\,$. Our analysis yields several hundred new integrable Hamiltonian families which also appear to be interesting from an algebro-geometric point of view. 

While ML has been applied to the analysis of integrable systems \cite{liu2022machine,liu2021machine,bondesan2019learning,Krippendorf:2021lee,Lal:2023dkj,LopesCardoso:2024tol} and, more broadly, to the study of symmetries in physical systems \cite{chen2023machine,Chen:2020jjw,
liu2022machine,liu2021machine,bondesan2019learning,melkosiamese,forestano2023deep}, this work is the first to systematically derive analytic expressions for multiple new Hamiltonians across various settings.
We now overview our framework with an example.
\section{Overview}
\label{sec:workflow}
The YBE is system of \(d^6\) cubic functional equations in \(d^4\) variables. As such it is non-linear and highly overdetermined, which makes even numerical algorithmic scans for solutions highly prohibitive.  In this paper, we introduce a new approach that combines a highly efficient neural network-driven numerical search with a procedure for extracting the analytical form of new integrable models. While a comprehensive explanation can only be made after the key ingredients are suitably defined, we will anticipate our results with an illustrative new model\,.
\begin{enumerate}
    \item \textit{Initialization:} We start by choosing a suitable ansatz, motivated either by generality, symmetry conditions or even the prior knowledge of an existing ansatz. In our illustrative case, we take \(d=3\) and consider R-matrices with subsets of activated vertices from this pattern:
\begin{gather} \label{SymPatternR}
R=\begin{pmatrix}
 * & 0 & * & 0 & * & 0 & * & 0 & * \\
 0 & * & 0 & * & 0 & * & 0 & * & 0 \\
 * & 0 & * & 0 & * & 0 & * & 0 & * \\
 0 & * & 0 & * & 0 & * & 0 & * & 0 \\
 * & 0 & * & 0 & * & 0 & * & 0 & * \\
 0 & * & 0 & * & 0 & * & 0 & * & 0 \\
 * & 0 & * & 0 & * & 0 & * & 0 & * \\
 0 & * & 0 & * & 0 & * & 0 & * & 0 \\
 * & 0 & * & 0 & * & 0 & * & 0 & *
 \end{pmatrix}\,,
\end{gather}
\item \textit{Exploration:} We use the neural network to scan this ansatz for the presence of new solutions of YBE, aided by the integrability criterion defined below in ~\eqref{eq:q2q3constraint}. 
Typically, the Hamiltonians thus discovered obey the integrability criterion ~\eqref{eq:q2q3constraint} to \(\mathcal{O}\left(10^{-4}\right)\). See Section ~\ref{sec:ML} for more details.
\item \textit{Hamiltonian Extraction:} We look for an algebraic variety of integrable models in the vicinity of the numerically found Hamiltonian. The general way to do this is somewhat involved and we postpone the discussion to Section \ref{sec:extractHamiltonian}. One point worth mentioning here is that the method for extracting exact relations relies on the integer nature of the coefficients in the polynomial relations between Hamiltonian entries, which enables rounding of numerical data. While we do not have a formal proof yet, our findings thus far suggest that deriving polynomial relations with integer coefficients has been sufficient for solving \ref{eq:q2q3constraint}.
One of the simplest integrable families we discovered from the \eqref{SymPatternR} pattern is a 25-vertex model that forms the following three-parameter linear variety \(H_{25}\):
\begin{equation}
\label{eq:hvariety}
\begin{split}
\alpha &= h_{ii}\,,\quad i=1\ldots 9\,,\\
\beta &= h_{15},\, h_{95},\, -h_{35},\, -h_{75},\,\\
&\quad -h_{5i}\,,\quad i = 1,\,3,\,7,\,9\,,\\
\gamma &= h_{24},\,h_{26},\, h_{62},\, h_{68},\\
&\quad -h_{42},\,-h_{48},\,
        -h_{84},\, -h_{86}\,.
\end{split}
\end{equation}
\item \textit{R-matrix:} The R-matrix is extracted using the Hamiltonian, derived in the previous step, as a starting point. We utilize one of the standard methods such as small-parameter expansion, or solving the Sutherland equations \cite{de2019classifying,deLeeuw:2020ahe,deLeeuw:2020xrw,Corcoran:2023zax}. In the present case, the final result obtained is remarkably simple. We find
\begin{equation}
    R\left(u\right) = P\exp\left(u\,H_{25}\right)\,,
\end{equation}
where \(P\) is a permutation matrix \eqref{eq:regularity} and \(\exp\) is the matrix exponential.
It can be explicitly checked that this R-matrix obeys the Yang-Baxter equation. R-matrices found for other new integrable systems are significantly more complicated. 
\end{enumerate}
\begin{figure}
   \includegraphics[scale=0.5]{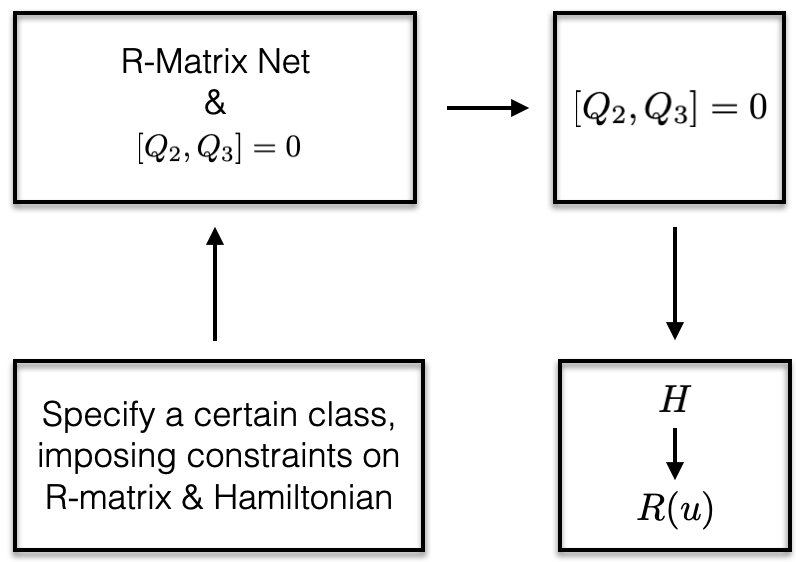}
   \label{fig:Scheme} 
\caption{Compact Scheme of our workflow}\label{fig:CompactScheme}
\end{figure}
The outlined workflow is schematically visualized in Figure \ref{fig:CompactScheme}.
\section{Quantum integrable spin-chains}
\label{sec:spinchainoverview}
We consider quantum-integrable spin chains with \(L\) site vector spaces \(V_i\)
isomorphic to \(\mathbb{C}^d\). 
The Hilbert space is then the 
$L$-fold tensor product \(\mathbb{V}=V_1\otimes ... \otimes V_L\).
The Hamiltonian \(H\) of a spin chain with nearest-neighbour interaction is a sum of two-site Hamiltonians \(H_{i,i+1}\) which encode interactions between degrees of freedom on the \(i\)-th and the \(i+1\)-th site respectively:
\begin{equation} \label{eq:SpinChainHamiltonian}
H=\sum_{i=1}^{L}H_{i,i+1}\,,
\end{equation}
where periodic boundary conditions  \(H_{L,L+1}\equiv H_{L,1}\) are assumed. 
Quantum integrability 
implies the existence of a tower of the mutually commuting charges \(\mathbb{Q}_n\):
\begin{equation}\label{eq:q2qnintro}
[\mathbb{Q}_m,\mathbb{Q}_n] =0\,.
\end{equation}
The construction of these charges and the proof of quantum integrability is based on
the existence of the R-matrix operator \(R_{ij}\in End\left(V_i\otimes V_j\right)\) which encodes the Hamiltonian
and higher conserved charges, as reviewed below. 
The R-matrix elements are generally bi-holomorphic functions \(R_{ij}\left(u,v\right)\) of two spectral parameters \(u,v\in\mathbb{C}\), but we specialize to the case of \textit{difference form} i.e.
\begin{equation}
    R_{ij}(u,v) \equiv R_{ij}(u-v)\,.
\end{equation}
These R-matrices are constrained to obey the Yang-Baxter equation \cite{Faddeev:1996iy,Perk:2006wiu}
\begin{equation}
\label{eq:ybequation}
R_{ij}(u-v)R_{ik}(u)R_{jk}(v)=R_{jk}(v)R_{ik}(u)R_{ij}(u-v)\,,
\end{equation}
where both the operators on the left-hand and right-hand sides are endomorphisms
over \(V_{i}\otimes V_j\otimes V_k\). 
In addition, imposing locality further fixes 
\begin{gather}\label{eq:regularity}
R_{ij}(0)=P_{ij}\,,
\end{gather}
where \(P_{ij}\) is the permutation matrix.
All charges \(\mathbb{Q}_{n}\) are encoded in the R-matrix through the transfer matrix \(T\left(u\right)\) defined as
\begin{equation}
    T(u)=\mathrm{tr}_a\left(R_{a,L}(u)R_{a,L-1}(u)\dots R_{a,1}(u)\right)\,,
\end{equation}
where \(a\) denotes an auxiliary spin site. The transfer matrix can be formally expanded to extract the conserved charges through
\begin{equation}
    \log{T(u)}=\sum_{n=0}^{\infty} \mathbb{Q}_{n+1} \frac{u^n}{n!}\,.
\end{equation}
In particular, the second charge is just the Hamiltonian \(\mathbb{Q}_{2}=H\) introduced earlier in equation \eqref{eq:SpinChainHamiltonian} and the  Hamiltonian density \(H_{i,i+1}\)  can be generated from the R-matrix using
\begin{equation}
\begin{split}
H_{i,i+1}&=R_{i,i+1}^{-1}(u)\frac{d}{du}R_{i,i+1}(u)|_{u=0}\\&=P_{i,i+1}\frac{d}{du}R_{i,i+1}(u)|_{u=0} \,,
\end{split}
\label{eq:HfromR}
\end{equation}
where $P_{i,i+1}$ is the permutation operator between sites $i$ and $i+1$. 
Let's mention that, given a solution of the Yang-Baxter equation, one can generate new ones by acting on the R-matrix with the following transformations :  i) similarity transformation : \((\Omega\otimes \Omega) R(u) (\Omega^{-1}\otimes \Omega^{-1})\) where \(\Omega\in GL(V)\) is a basis transformation, ii) rescaling of the spectral parameter : \(u\rightarrow c\, u\,, \ \forall\, c\in \mathbb{C}\), iii) multiplication by any scalar holomorphic function \(f(u)\) preserving regularity condition  : \(R(u)\rightarrow f(u)R(u)\), \(f(0)=1\). This degree of freedom can be used to set one of the entries of \(R\)-matrix to one or any other fixed function, iv) permutation, transposition and their composition: \(PR(u)P,\ R(u)^T,\ PR^T(u)P\). It means that one is rather interested in the classes of equivalence modulo these transformations. However, in what follows we will not aim at the full classification problem and the full fixing of all these degrees of freedom will not be crucial for us.

\section{Families of integrable models as algebraic varieties}
\label{sec:reshetikhin}
As was shown in \cite{Tetelman:1982}, higher charges can be generated with the use of the Boost operator \(\mathcal{B}\) :
\begin{gather}
\mathcal{B}=\sum\limits_{a=-\infty}^\infty a H_{a,a+1}
\end{gather}
as  commutators:
\begin{gather}
\mathbb{Q}_{r+1}=[\mathcal{B},\mathbb{Q}_{r}]\,.
\label{HigherChargesWithBoost}
\end{gather}
For spin-chains of finite length $L$, these charges can be written as a sum over the \(r\)-site Hamiltonian densities \(H_{a,...,a+r}\) as follows:
\begin{gather}
\mathbb{Q}^L_{r}=\sum\limits_{a=1}^L H_{a,a+1,..,a+r-1} \,.\label{HigherChargesFiniteL} 
\end{gather}
For example, the density of \(\mathbb{Q}_3\) is given by the following commutator :
\begin{gather}
H_{a,a+1,a+2}=[H_{a,a+1},H_{a+1,a+2}]\,.
\end{gather}
In the following, we shall focus on the constraint 
\begin{equation}\label{eq:q2q3constraint}
[\mathbb{Q}^L_2,\mathbb{Q}^L_3] = 0\,,
\end{equation}
This is a necessary condition for quantum integrability, that has long been conjectured to be a sufficient one as well \cite{Grabowski:1994rb}. It can also been seen as the global version of the \textit{Reshetikhin condition}. To the best of our knowledge, there is neither a proof, nor any counterexamples. Formally, a solution of \eqref{eq:q2q3constraint} is just a potentially integrable Hamiltonian. However, given the strong evidence in favor of this conjecture, we will just write ``integrable" instead of ``potentially integrable". In order to prove integrability, one should obtain the R-matrix associated with the Hamiltonian satisfying the above Reshetikhin condition, and we will illustrate several explicit examples for the same. However, the most nontrivial part is identification of the integrable Hamiltonian, and it will be the main focus for us. In order for this condition \eqref{eq:q2q3constraint} to be non-degenerate, the minimal length of the spin-chain should be \(L=4\). If the site-space is \(d\)-dimensional, the corresponding two-particle Hamiltonian has \(d^{4}\) parameters
while the constraint \eqref{eq:q2q3constraint} furnishes \(d^{8}\) scalar equations.  For example in the case of \(d=3\) the condition \eqref{eq:q2q3constraint} thus imposes \(81\times 81 = 6561\)
cubic polynomial equations on the \(9\times 9 = 81\) entries of the Hamiltonian density - complexity inaccessible for any existing computational methods.  
Using \eqref{HigherChargesFiniteL} and \eqref{eq:q2q3constraint} one can see that the condition \eqref{eq:q2q3constraint} provides the system of homogeneous cubic equations with integer coefficients on the entries of the Hamiltonian density :
\begin{gather}
H=H_{i,i+1}=\begin{pmatrix}
h_{11}  & ...  & h_{1d^2}\\
... & ...& ... \\
h_{d^21} & ... & h_{d^2d^2}
\end{pmatrix}\,,
\end{gather}
These equations define an algebraic variety in \(\mathbb{CP}^{d^4-1}\) whose irreducible components are families of integrable models. Namely, let's introduce the ideal \(I\subset\mathbb{C}[h_{ij}]\) generated by this system of equations, then the algebraic variety \(V(I)\) can be decomposed into irreducible components 
\begin{gather}\label{IrredComponents}
V(I)=\bigcup_i V(I_i)\,,
\end{gather}
where each \(I_i\) is a prime ideal. The fact that the coefficients in \eqref{eq:q2q3constraint} are integers does not guarantee that the system of defining equations for irreducible components can also be chosen with all integer coefficients. However, as we will see in all our examples, extracting polynomial relations with integer coefficients turnes out to be sufficient to identify the irreducible component.

The groups of transformations i)-iv) mentioned above act on the irreducible components, and the factors are again algebraic varieties. In this paper our prime goal is to set up the framework for the  discovering new integrable models in various  classes, rather than full classification so we will not care about the global parametrization of this space of orbits and instead will use the standard in physics literature notation of vertex models, specifying nonzero components of the Hamiltonian density and R-matrix. We can use a gauge transformation \(\Omega\in SL(d,\mathbb{C}) \) to set \(d^2-1\) elements of the \(d^4\) entries of \(H\) to zero, which practically fixes the gauge. In what follows, we usually will consider the \(k\)-vertex models with \(k< d^4-d^2+1\). We will specify a certain ``pattern" - set of nonzero entries of R-matrix as \(\Pi_R\) and the corresponding pattern for the Hamiltonian as \(\Pi_H\). Fixing the gauge or pattern reduces the number of independent variables and equations, but the problem remains highly complex. As sketched in the introduction, we will utilize neural networks to overcome it. 
\section{Neural Networks as numerical solvers of the Yang-Baxter Equation} \label{sec:ML}
We will utilize neural networks to find numerical solutions of YBE, which will subsequently serve as a seed for further numerical refinement and extraction of analytical expressions. At a conceptual level, neural networks allow us the flexibility of scanning across the vast variety of candidate solution spaces relatively quickly and also deal with the non-convexity of the optimization problem at hand. We had previously proposed the architecture \texttt{R-Matrix Net} for precisely this kind of analysis \cite{Lal:2023dkj}. 
The building block for the \texttt{R-Matrix Net} is the fully-connected neural network, i.e. the \textit{multi-layer perceptron} (MLP). These networks consist of
an input layer \(a^{in}\in \mathbb{R}^{n_0}\), followed by a series of fully connected layers 
and terminate in an output layer \(a^{out}\in \mathbb{R}^{n_{L+1}}\). 
More explicitly it is an ansatz which has the form of the composition of alternated affine and nonlinear transformations: 
\begin{gather}\label{NNcomposition}
a^{out}=\tilde{h} \circ A^{(L+1)} \circ h \circ A^{(L)} \circ ... \circ h \circ A^{(1)} \circ a^{in}\,,
\end{gather}
where affine transform \(A^{(\ell)}(a^{(\ell-1)})\equiv w^{(\ell)} a^{(\ell-1)}+b^{(\ell)}\) is parametrized by a \textit{weight matrix} $w^{(l)}\in \mathcal{M}(n_{l},n_{l-1},\mathbb{R})$ and \textit{bias vector} - $b^{(l)}\in \mathbb{R}^{n_l}$ . \(h\) is a non-linear, non-polynomial activation function which acts element-wise over \(z\) while \(\tilde{h}\) is the activation function for the \textit{output layer}, which is just the identity map in our case. These MLPs are trained by optimizing a loss function in terms of parameters \(w^\ell\) and \(b^\ell\). In the most basic implementation, this loss function may encode pairs of input-output values, but it is well established by now that loss functions can also efficiently encode more complex properties of the target function such as being solutions of differential or functional equations. The feasibility of this approach is motivated by the well known universal approximation property of MLPs \cite{cybenko1989approximation,hornik1989multilayer,lu2017expressive,hoffman2019robust,park2021minimum} along with their feature learning capabilities. Together, these provide justification that a neural network may approximate any function over a finite interval of arguments and the desired properties of the target function be encoded in loss functions with respect to which the neural network can be trained. These observations are also the key driver to the automated search for R-matrices which
we have implemented here and previously in \cite{Lal:2023dkj}. It was proposed that each non-zero function in the R-matrix ansatz be modeled by such a neural network and the overall architecture be trained on loss functions encoding the Yang-Baxter equation, locality and additional constraints on the Hamiltonian and/or R-matrix, if any. In particular, we define the matrix norm \(||...||\) as 
\begin{equation}
    ||A||= \sum\limits_{\alpha,\beta=1}^n \vert A_{\alpha\beta}\vert 
\end{equation}
for a complex-valued \(n\times n\)  matrix \(A\) then the
Yang-Baxter loss
\begin{equation}
\label{eq:lossybe}
\begin{split}
\mathcal{L}_{YBE} = 
    \vert\vert 
    &\mathcal{R}_{12}(u_c)\mathcal{R}_{13}(u_a)\mathcal{R}_{23}(u_b)
    \\&-\mathcal{R}_{23}(u_b)\mathcal{R}_{13}(u_a)\mathcal{R}_{12}(u_c)\vert\vert
    \,,
\end{split}
\end{equation}
where \(\{u_a,u_b,u_c\equiv u_a-u_b\}\) are points drawn from the interval \((-1,1)\).
The locality condition ~\eqref{eq:regularity} is encoded in the loss functions as 
\begin{gather}\label{eq:regularityloss}
    \mathcal{L}_{reg} = || \mathcal{R}\left(0\right) -P
    ||.
\end{gather}
In the similar manner we can implement other constraints like hermiticity of the Hamiltonian, braiding unitarity, crossing etc. These losses are positive semi-definite and vanish only when the corresponding functional identities are exactly satisfied. Finally once we found a certain Hamiltonian from the given class we can activate repulsion loss 
\begin{equation}\label{eq:replusionloss}
\mathcal{L}_{repulsion}=\exp{(-||H-H_o||/\sigma)}\,,
\end{equation}
for several epochs to scan the vicinity of the found Hamiltonian for other representatives from the same class, see \cite{Lal:2023dkj} for further details.  
In this paper we will make the main focus on the most nontrivial YBE loss assuming a certain pattern for the R-matrix and locality, no other restrictions will be imposed. In order to navigate the the search we will also implement the loss function \(\mathcal{L}_{Q_2Q_3}\) encoding Reshetikhin condition :
\begin{equation}\label{eq:lossq2q3}
\mathcal{L}_{Q_2Q_3} =
  \max \vert [\mathbb{Q}_2,\mathbb{Q}_3]\vert
    \,.
\end{equation}
where   \(\max \vert M \vert =\max\limits_{i,j} |M_{ij}|\). This loss not only improves the training but also provides a natural metric to measure the ``closeness" of the found Hamiltonian to an integrable one. 
Empirically, we observed that a terminal value of \(\mathcal{L}_{Q_2Q_3} \sim \mathcal{O}\left(10^{-4}\right)\) corresponds to a candidate integrable Hamiltonian family which can be found by our subsequent methods.
\subsection{Averting Mode Collapse} In order to prevent the neural network from converging onto a trivial solution, a phenomenon known as \textit{mode collapse}, we may introduce a loss function that penalizes such solutions. More generally we can prevent the convergence to already known solutions. For example, to encourage \(\mathcal{O}\left(1\right)\) terms in the Hamiltonian, we may introduce a loss function ensuring unit norm for the weighted absolute sum of the Hamiltonian entries $H_{ij}$, i.e.
\begin{equation}
    \mathcal{L}_{m.c.} = \left| \left( \frac{1}{n} \sum_{i,j} \left| H_{ij} \right| \right) - 1 \right|\,,
\end{equation}
$n$ being the number of non-zero Hamiltonian entries. 
One natural extension of this loss is to differently penalize the vanishing of different Hamiltonian entries. For example, 
\begin{equation}
\begin{split}
    \tilde{\mathcal{L}}_{m.c.} =& \left| \left( \frac{1}{n_{diag}} \sum_{i} \left| H_{ii} \right| \right) - \lambda_{diag} \right|\\
    &+
    \left| \left( \frac{1}{n_{off-diag}} \sum_{i\neq j} \left| H_{ij} \right| \right) - \lambda_{off-diag} \right|\,,
\end{split}
\end{equation}
where we distinguish between diagonal and off-diagonal entries in the learnt Hamiltonian, and $\lambda_{diag},\lambda_{off-diag}$ are the hyperparameters which control the relative strengths of the two repulsions. As a further extension, if we want to avoid solution-classes satisfying known constraints, say $\{C_k\}$ we can add a further term to this mode collapse,
\begin{equation}\label{mc-loss-general}
\begin{split}
    \mathcal{L}^{total}_{m.c.} = & \tilde{\mathcal{L}}_{m.c.}+ \left| \left(  \sum_{k} \mu_k\left| C_{k} \right| \right) - \lambda_{repul} \right|\,,
\end{split}
\end{equation}
where $\mu_{k},\lambda_{repul}$ are further new hyperparameters. In practice, we fix these hyperparameters randomly to some positive values between 0 and 1 at the start of each optimization run. For example if we want to repulse from the given family of Hamiltonians we can choose \(\{C_k\}\) as a Gr{\"o}bner basis \(G_k(\{h_{ij}\})\) of the ideal defining this subvariety and set \(\lambda_{repul}=1\), \(\mu_k=1\).
\subsection{Neural Network Implementation and Training}
\begin{figure}
   \includegraphics[scale=0.33]{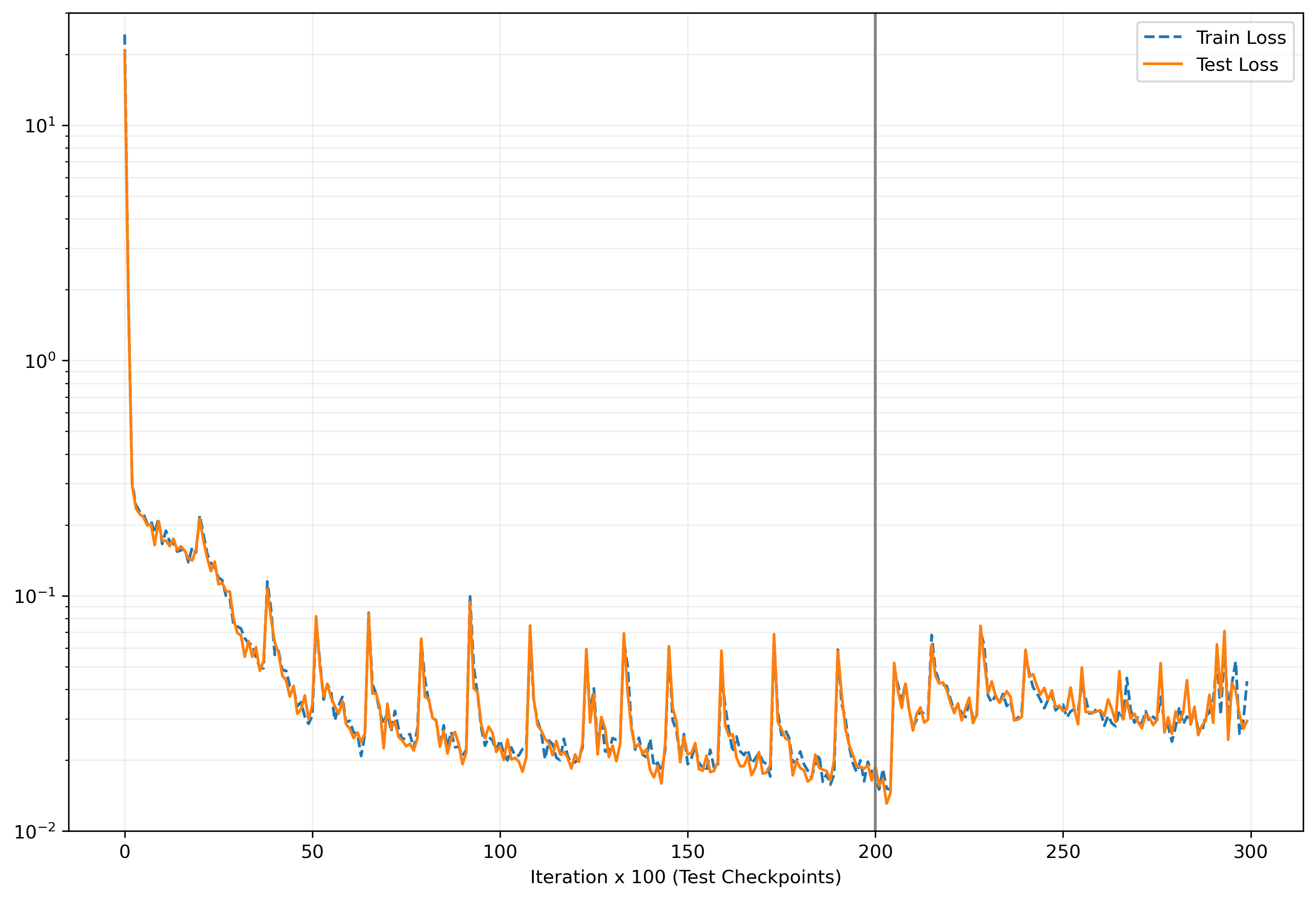}
\caption{The evolution of the training and validation losses for a typical round of training the neural network. The x axis is plotted in units of 100 iterations of gradient descent. The y axis is in log scale. The mode collapse loss is turned off at \(200\times 100\) iterations, indicated by the gray vertical line.}\label{fig:train_val_evol}
\end{figure}

\begin{figure}
   \includegraphics[scale=0.33]{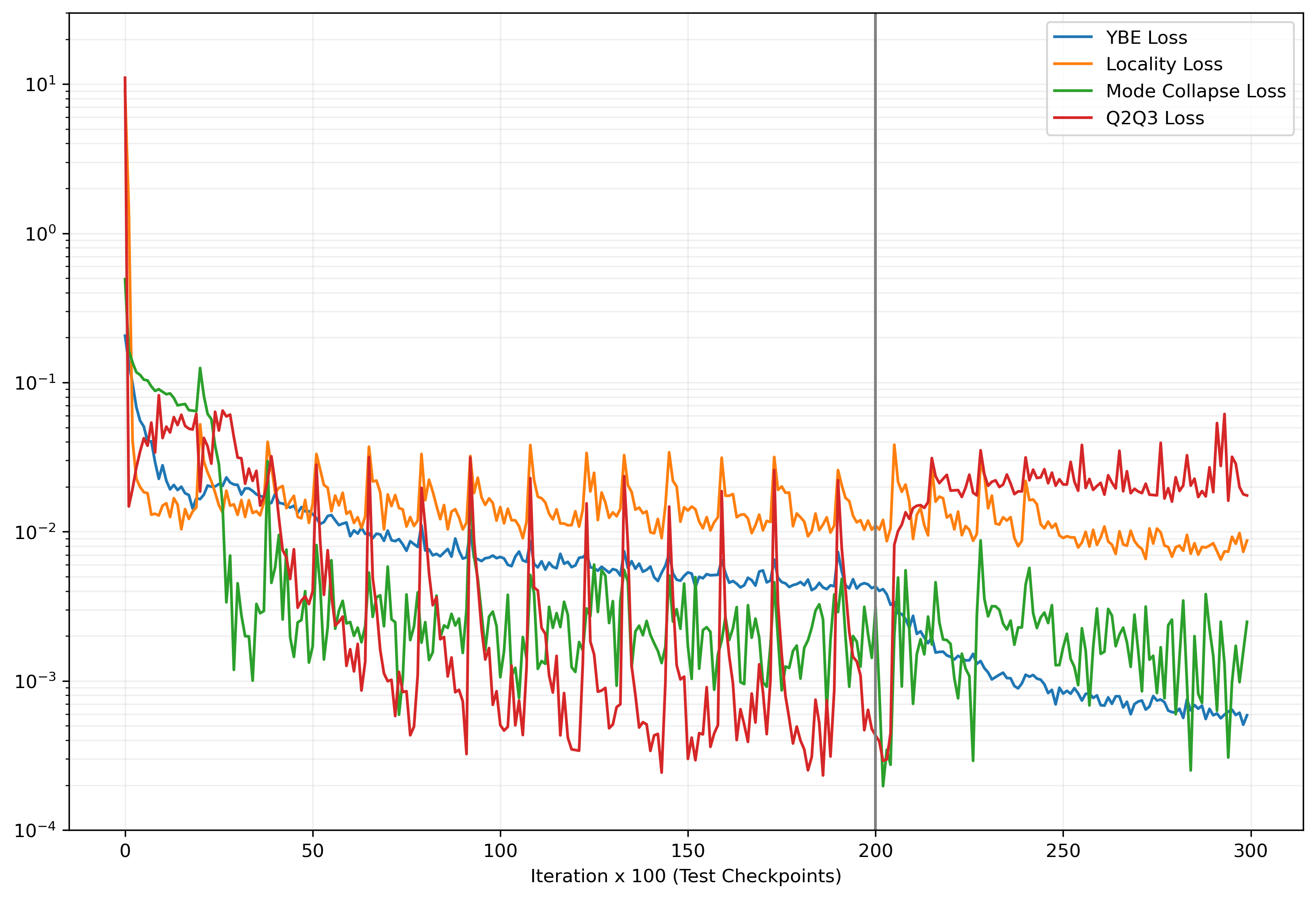}
\caption{The decomposition of the validation loss into its various constituents in a typical round of training the network. The x axis is plotted in units of 100 iterations of gradient descent. The y axis is in log scale. The mode collapse loss is turned off at \(200\times 100\) iterations, indicated by the gray vertical line.}\label{fig:loss_decomp_evol}
\end{figure}

We now overview the neural network architecture and the hyperparameters for training the 
\texttt{R-Matrix Net} which we introduced previously in \cite{Lal:2023dkj}. The architecture is a set of independent MLPs, in one-to-one correspondence with the non-vanishing entries of the R-matrix dictated by the pattern. In principle, this would yield an architecture comprising \(\mathcal{O}\left(d^4\right)\) number of independent neural networks which are trained using the loss function
\begin{equation}\label{eq:totalloss}
    \mathcal{L} =\sum_{\Lambda}w_\Lambda \mathcal{L}_\Lambda\,,
\end{equation}
where \(w_\Lambda\) is the relative weight for the loss term \(\mathcal{L}_\Lambda\) and 
$\Lambda$=$\{YBE\,, reg\,, m.c.\,, Q_2Q_3\}$.
Given the computational demands, we rewrote the \texttt{R-Matrix Net} from the ground up in \texttt{JAX} which utilizes just-in-time (JIT) compilation and additionally provides highly sophisticated auto-differentiation capabilities which are also crucial for implementing our loss functions \cite{jax2018github}. 
The hyperparameters \(w_\Lambda\) and others are further elaborated on in Table \ref{tab:hyperparameters}. Training typically proceeds for $\mathcal{N}_{steps}=50000$ iterations of gradient descent and a batch size of 32 to 128 \(u,v\) pairs sampled randomly from \(\Omega = \left(-1,1\right)\) at each iteration. Default run involves attenuating the weights $w_{m.c.}, w_{Q_2Q_3}$ by a factor of 10 after a fixed number of iterations, we call the turn-off time $t_{TO}$, to allow YBE loss and regularity to drive the final phase of optimization. The evolution of various losses during typical rounds of training is shown in Figures \ref{fig:train_val_evol} and \ref{fig:loss_decomp_evol}.
\begin{table}
\caption{The hyperparameters of \texttt{R-Matrix Net}, representative choices for their values and scheduling policy, as applicable.}
\label{tab:hyperparameters}
\begin{ruledtabular}
\begin{tabular}{lp{5cm}}
\textbf{Parameter} & \textbf{Configuration} \\ \hline
\(\eta\) (Adam learning rate) & 
  \begin{tabular}{l}
  Initial value: \(10^{-3}\) \\
  Final value: \(10^{-8}\) \\
  Scheduling: annealing factor, \\
  monitor val. loss, \\
  patience = 500 iterations
  \end{tabular} \\ \hline
\(w_{YBE}\) & 
  \begin{tabular}{l}
  Weight for \(\mathcal{L}_{YBE}\) in net loss \\
  Value: 1.0 (constant)
  \end{tabular} \\ \hline
\(w_{reg}\) & 
  \begin{tabular}{l}
  Weight for \(\mathcal{L}_{reg}\) in net loss \\
  Value: 1.0 (constant)
  \end{tabular} \\ \hline
\(w_{m.c.}\) & 
  \begin{tabular}{l}
  Weight for \(\mathcal{L}_{m.c.}\) in net loss \\
  Penalizes trivial/known solutions \\
  Init value: 1.0 (constant) \\
  Final value: 0.1(after\\ $t_{TO}$=20,000 iterations)
  \end{tabular} \\ \hline
\(w_{Q_2Q_3}\) & 
  \begin{tabular}{l}
  Weight for \(\mathcal{L}_{Q_2Q_3}\) in net loss \\
  Sub-leading to $\mathcal{L}_{m.c.}$ \\
  Init value: 0.1 (constant)\\
  Final value: 0.01 (after \\$t_{TO}$=20,000 iterations)
  \end{tabular} \\
\end{tabular}
\end{ruledtabular}
\end{table}
\section{Hamiltonian Extraction}
\label{sec:extractHamiltonian}
In principle, the primary decomposition \eqref{IrredComponents} can be performed using Gr\"{o}bner basis or other standard computational techniques. However, the complexity of these algorithms typically increases very rapidly—often double exponentially—making them impractical for the cases of interest. Indeed even in the simplest case of two-dimensional site-space we have 256 cubic equations \eqref{eq:q2q3constraint} depending on 16 variables. However, as one can see in \cite{de2019classifying}, the Hamiltonians are extremely simple, there are just 14 different classes with 12 families forming linear varieties and two others involving simple quadratic relations. Other examples from literature demonstrate similar simplicity, suggesting that we should look for such relations between Hamiltonian entries and then use them to solve the system \eqref{eq:q2q3constraint}. Let's describe this strategy in more details.

1) \texttt{R-Matrix Net} outputs a new numerical Hamiltonian \(H_{seed}\) with entries of order \(O(1)\) and typical precision \(\sim 10^{-4}-10^{-3}\).

2) We next solve algebraic equations  \eqref{eq:q2q3constraint} in the vicinity of \(H_{seed}\). At the given precision, the problem turns out to be (almost) convex and standard numerical methods can improve precision to \(10^{-10}-10^{-8}\) or even higher if needed. We then perform \(N\) random perturbations \(\delta H\sim O(10^{-2})\) of the original Hamiltonian \(H_{seed}+\delta H\) and again solve \eqref{eq:q2q3constraint}. It gives us a cloud \(\mathcal{V}\) of high-precision Hamiltonians \(\mathcal{V}=\{H_{num}^{(k)}\}\), \(1\leq k\leq N\).

3) Because the perturbations \(\delta H\) are small, we expect that the Hamiltonians \(\mathcal{V}\) belong to a certain irreducible component \eqref{IrredComponents}, and as explained above, we expect various simple polynomial relations between nonzero Hamiltonian entries \(\{h_{ij}\}\) on \(\mathcal{V}\). Let's describe it in more details. First we consider just linear relations between \(K\) nonzero Hamiltonian entries \(\eta=\{h_{ij}\}\) and we will use a single index \(\alpha\in (1,..,K)\) to numerate them. As an initial step, we construct \(N\) lists \(\eta^{(k)}\) from \(H_{num}^{(k)}\) and combine them into the one \(N\times K\) matrix \(M=\{\eta^{(1)},..,\eta^{(N)}\}\). Because the numerical Hamiltonians \(\mathcal{V}\) are generated from random perturbations, they will all be in general positions, and we can choose \(N=K\). In order to find the linear relations on \(\mathcal{V}\) we use the singular value decomposition \((u,\sigma,v)\) :
\begin{gather}
M=u\sigma v^\dag\,,
\end{gather}
where \(\sigma\) with singular values of \(M\) on the diagonal ordered as \(\sigma_{i,i}\geq\sigma_{i+1,i+1}\). Vanishing singular values \(\sigma_{i,i}=0\), \(i\geq i_0\) correspond to linear relations :
\begin{gather}
R_i=\sum\limits_{\alpha=1}^K v_{\alpha,i}\eta_\alpha=0,\ \ \ i\geq i_0\,.
\end{gather}
Let's mention that in practice vanishing singular values are not exactly zeros but small numbers and we introduce an cutoff to distinguish them from non-vanishing ones.  
Finally we choose \(i_0-1\) independent variables (or less if the rank is smaller) , let's say \(\eta_1,...,\eta_{i_0-1}\) and solve the system 
\begin{gather}
R_i=0, \ \ \ i\geq i_0\,,
\end{gather}
with respect to them:
\begin{gather}
\eta_\beta=\sum\limits_{i=1}^{i_0-1}\tilde{c}_{\beta,i}\eta_i, \ \ \ \beta>i_0-1\,.
\end{gather}
In all cases we analyzed, the coefficients \(\tilde{c}_{\beta,i}\) were very close to either integers or rational numbers with small numerator and denominator if the precision for \(\mathcal{V}\) was high enough, which suggests that we should just round them \(c_{\beta,i}=Round_\mathbb{Q}[\tilde{c}_{\beta,i}]\) to the "closest" rational number and use 
\begin{gather}\label{LinRelations}
\eta_\beta=\sum\limits_{i=1}^{i_0-1}c_{\beta,i}\eta_i
\end{gather}
as exact relations.
In principle one can use them to reduce the number of independent Hamiltonian entries and then repeat the similar algorithm to extract the linear relations between more general monomials \(\{\eta_i\},\{\eta_i\eta_j\},...,\) however, in practice even linear relations \eqref{LinRelations} often turn out to be enough to drastically simplify and solve the system \eqref{eq:q2q3constraint}.

The crucial element of this procedure is the rationality of coefficients \(c_{\beta,i}\). Although a formal proof is not yet available, all cases studied so far indicate that extracting polynomial relations with integer coefficients\footnote{after the multiplication by the common denominator we can speak just about integer numbers instead of rational} has been sufficient to obtain a solution.

\section{Examples of New Models}
\label{sec:results}
The ingredients described in Sections \ref{sec:ML} and \ref{sec:extractHamiltonian} yield a few hundred new Hamiltonians that are integrable in the sense of the Reshetikhin criterion \eqref{eq:q2q3constraint}, too numerous to exhaustively detail here. We will therefore highlight only a few highly illustrative examples, and provide R-matrices for some of them. The discussion that follows should be read in conjunction with the workflow outlined in Section \ref{sec:workflow}. 
\paragraph{Example 1} The Hamiltonian we presented in Section \ref{sec:workflow}, see Equation \eqref{eq:hvariety}.
\paragraph{Example 2} Lets look at another general pattern that likely encompasses many families of integrable Hamiltonians. In \cite{de2019classifying} authors found six classes of integrable spin-chains with two-dimensional site space and upper triangular R-matrix (plus terms in lower triangular part originated from permutation).  Generalizing such an ansatz to three dimensions, we start with an upper triangular R-matrix.

\begin{gather}\label{eq:uppertr_ansatz}
R=\begin{pmatrix}
 * & * & * & * & * & * & * & * & * \\
 0 & * & * & * & * & * & * & * & * \\
 0 & 0 & * & * & * & * & * & * & * \\
 0 & * & 0 & * & * & * & * & * & * \\
 0 & 0 & 0 & 0 & * & * & * & * & * \\
 0 & 0 & 0 & 0 & 0 & * & * & * & * \\
 0 & 0 & * & 0 & 0 & 0 & * & * & * \\
 0 & 0 & 0 & 0 & 0 & * & 0 & * & * \\
 0 & 0 & 0 & 0 & 0 & 0 & 0 & 0 & *
 \end{pmatrix}\,,
\end{gather}

One of the simplest models we found is the following 15-vertex model consisting of 9 diagonal terms \(h_{ii}\) and six off-diagonal \(h_{13},h_{17},h_{19},h_{39},h_{46},h_{79}\) with the following relations :
\begin{equation}
\label{eq:uppertriangularsol}
\begin{split}
 h_{11}=h_{99},\,h_{66}=h_{55}\,,\\ h_{33}=h_{11}-h_{44}+h_{55},\ h_{77}=h_{11}+h_{44}-h_{55}\,,\\
 h_{88}=h_{22}+h_{44}-h_{55}, \ h_{39}=h_{13}+h_{17}-h_{79}\,.
\end{split}
\end{equation}
The corresponding R-matrix is somewhat lengthy and is given below in Equations \eqref{eq:uppertr_rmatrix1} and
\eqref{eq:uppertr_rmatrix2}. 

In appendix~\eqref{app:uppertr_hamil_19v} we list another pair of new upper-triangular integrable families with 19 vertex Hamiltonians, which differ only by \(h_{55}\), illustrating how close the models can be. Many other examples of models obtained from the upper-triangular ansatz can be found in the attached file.

\paragraph{Example 3} Another strategy is to relax (symmetry) constraints in known models. For example, we can consider the general 19 vertex ice-rule ansatz (see equation~\eqref{eq:ice-rule}) and relax any additional symmetry assumptions (see \cite{idzumi1994solvable}). We found several interesting solutions in the setting, such as the following two-parameter family of Hamiltonians:
\begin{equation}\label{ice-rule-example}
\begin{split}
h_{11} = h_{99} = 1\,,\\
h_{24}=h_{42}=h_{37}=h_{73}=h_{68}=h_{86}=\alpha\,,\\
h_{35}=h_{53}=h_{57}=h_{75} = \beta\,,\\
h_{22}=h_{66}=-1 +\alpha -\tfrac{\beta^2}{\alpha}\,,\\
h_{33}=-h_{44}=-h_{88}=-3+3\alpha-\tfrac{2\beta^2}{\alpha}\,,\\
h_{55}=1-2\alpha+\tfrac{\beta^2}{\alpha},\,\,
h_{77}=5-5\alpha+\tfrac{4\beta^2}{\alpha}\,.
\end{split}
\end{equation}
Another such example is presented in Appendix~\eqref{app:ice-rule-example2}.
\paragraph{Example 4} is a \(d=4\) spin system obtained from an extension of the \(d=4\) XXZ pattern (see \cite{de2020new} for an earlier analytic search). 
The 2-site Hamiltonian $H_{IJ}$ is a \(16\times 16\) matrix, with indices $I,J$ enumerated using the hexadecimal system \{1,\,\ldots,\,9,\,A,\,\ldots\,,G\}. The nonzero entries of the Hamiltonian include: all 16 diagonal terms \(h_{ii}\), plus 8 off-diagonal terms \(h_{25}\), \(h_{52}\), \(h_{4D}\), \(h_{D4}\),  \(h_{7A}\), \(h_{A7}\), \(h_{CF}\), \(h_{FC}\). These entries are related by the following identities
\begin{equation}
\label{eq:4dsol}
\begin{split}
&h_{11} = -h_{55} + h_{77}+h_{99},\,
h_{22} = h_{AA}-h_{55}+h_{77},\,\\ 
&h_{33} = -2h_{55}+2h_{77}+h_{99},\,
h_{66} = h_{AA}+h_{55}-h_{99},\,\\
&h_{BB} = -h_{55}+h_{77}+h_{99},\,
h_{CC} = h_{AA}-h_{FF}+h_{77},\,\\
&h_{DD}=h_{FF}+h_{55}-h_{77},\,
h_{GG} = h_{AA}+h_{55}-h_{99},\,\\
&h_{52} =\tfrac{h_{D4}h_{4D}}{h_{25}},\,h_{A7} = \tfrac{h_{D4}h_{4D}}{h_{7A}},\, h_{FC} = \tfrac{h_{D4}h_{4D}}{h_{CF}}\,, \\
&h_{44} = h_{AA}-h_{FF}-h_{55}+2h_{77}\,,\\
&h_{88} = h_{AA}-h_{FF}+h_{55}+h_{77}-h_{99}\,,\\
&h_{EE}= h_{AA}+h_{FF}+h_{55}-h_{77}-h_{99}\,.
\end{split} 
\end{equation}
\paragraph{Hundreds of other examples:}
For illustrative purposes, we attach a Mathematica file containing hundreds of additional examples. While this represents only a fraction of our findings, it provides insight into the diversity of models, even when they share the same vertices. As previously mentioned, new models can be generated by relaxing one or two constraints in already discovered ones, which is how some of the presented models were obtained. 

\section{Discussion and future directions}
In this paper we introduced a novel AI-based framework to discover new integrable models. It combines the \texttt{R-Matrix Net} neural network with the auxiliary system of algebraic equations \eqref{eq:q2q3constraint} and the procedure \ref{sec:extractHamiltonian} to extract exact relations defining Hamiltonian families from the deep-learned data.

From the algebraic geometry point of view we construct irreducible components of the algebraic variety defined by \eqref{eq:q2q3constraint}. The extraction of exact relations described in section \ref{sec:extractHamiltonian} crucially relies on the fact that coefficients are integers, which allows us to round off our numerical data. While we do not yet have a formal proof, all our cases so far have demonstrated that extracting polynomial relations with integer coefficients has been sufficient to solve  \ref{eq:q2q3constraint}. Moreover all integrable Hamiltonians we found, remarkably, can be written as rational varieties and we would like to conjecture that \textit{all irreducible components \eqref{IrredComponents} are (uni)rational varieties}. It is possible in principle, that the conjecture may eventually need to be qualified by an as yet unknown, but fairly general extra constraint. Reshetikhin condition has the form of the vanishing commutator, and in the simpler case of commuting varieties the rationality was already proved \cite{popov2008irregularsingularlocicommuting} \footnote{We thank Samuel Stark for bringing this reference to our attention.}. 
Some of the discovered Hamiltonians form highly complex varieties -- see e.g. \eqref{eq:bigvariety_b} -- making it tempting to initiate their systematic analysis using algebraic geometry. In particular, studying their topology by computing homology groups could reveal distinctive properties of this class of varieties and  inspire a new research direction in the algebro-geometric analysis of integrable models. 

Furthermore, by fully automating Hamiltonian extraction, we could potentially generate hundreds of thousands or even millions of new models, creating a novel type of big data. This dataset could, in turn, be analyzed using AI techniques, akin to the machine-learning approaches applied to the string landscape \cite{He:2017aed,ruehle2017evolving,krefl2017machine,Carifio:2017bov}.

We emphasize that the primary focus of this work was the discovery of integrable Hamiltonians, understood as solutions to the Reshetikhin condition. As mentioned in the introduction, all solutions found so far have led to corresponding R-matrices. However, in principle, verifying the existence of an associated R-matrix requires a separate analysis for each case. While we explicitly carried out this verification for only a few examples, it is important to highlight that our method is based on numerically solving the YBE and the high precision of these numerical solutions provides additional confirmation of the existence of R-matrices behind the extracted Hamiltonians.

The framework developed in this work is highly flexible and can be adapted to various integrable models beyond spin chains. In particular, it can be used to systematically search for new 2D integrable quantum field theories \cite{Zamolodchikov:1977nu}. In this case, additional constraints—such as unitarity, crossing symmetry, global symmetries, and analyticity—must be incorporated. These can be implemented directly at the neural network architecture level and through additional terms in the loss function.

Next, to set up the search for integrable string worldsheet sigma models on an AdS background, we need to extend our approach to S-matrices of non-difference form \cite{Arutyunov:2009ga,lloyd2015complete,frolov2022massless,cavaglia2021quantum}. In this case, the Boost operator will be modified by an additional derivative term with respect to the second spectral parameter, resulting into the Reshetikhin condition in the form of  first-order ODE system. However, we can formally treat these first derivatives as a new set of algebraic variables and analyze the resulting augmented algebraic variety. Once its irreducible components are extracted, the drastically simplified ODE system may become much easier to solve.

With mild modifications we also can  discover new Lindbladian systems \cite{de2021constructing,de2022bethe,de2024hidden,paletta2023yang}, spin-chains with medium-range interaction \cite{Gombor:2021nhn} and R-matrices providing quantum gates for quantum computers \cite{zhang2024optimal,singh2024unitary,ge2016yang,yu2014factorized,wang2009entanglement,chen2007braiding}. 

In addition, integrable models often exhibit symmetries associated with certain algebras, such as Yangian, Hecke, Temperley-Lieb algebras etc. The discovery of new integrable systems and their corresponding R-matrices is likely to uncover novel algebraic structures, which may be of independent interest to mathematicians. Further, R-matrices are well known as a source of topological knot invariants. Adapting our method to search for new invariants would be an exciting direction. In particular, for a given pair of knots that remain indistinguishable by known invariants, one could employ \texttt{R-Matrix Net} to search over R-matrices to identify new invariants aiming to distinguish the pair. This approach could be combined with other ML-based techniques to the analysis of knots \cite{gukov2023searching,gukov2024rigor,Gukov:2020qaj}.

The ability to extract exact analytical formulas places our project among the very few existing examples of AI-generated or AI-assisted mathematical discoveries.  In the medium- to long-term, our framework could be of interest not only to human researchers but also to AI agents (tool-use), serving as a testbed for automated scientific discovery. One can envision for instance an AI agent \cite{jansen2024discoveryworld, baek2024researchagent, romera2024mathematical, schmidgall2025agent, kumarappan2024leanagent,nathani2025mlgym}, built on a fine-tuned LLM, that analyzes arXiv papers to identify promising classes of models and then utilizes our framework to conduct an exact search for new integrable systems.

\section*{Acknowledgments}
SL is supported by Beijing Natural Science Foundation grant no IS23010. SM would like to thank the BLH program organized by INI, Cambridge, where part of the initial phase of the work was done. ES would like to thank Alexander Esterov for highlighting the non-triviality of the fact that all our Hamiltonian families form (uni)rational varieties. Computational grant for this work was provided by Nebius AI Cloud.
\par\medskip
\appendix
\section*{Appendix}
\section{R-matrix for the upper-triangular 15-vertex Model}
\label{app:uppertr_rmat}
We present the non-vanishing elements of the R-matrix corresponding to the 15-vertex Hamiltonian of Equation \eqref{eq:uppertriangularsol}.
\begin{equation}
\label{eq:uppertr_rmatrix1}
\begin{split}
r_{11} &= r_{99} = e^{h_{11}u}\,,\,r_{42} = e^{h_{22}u}\,,\, r_{24} = e^{h_{44}u}\,,\\  
r_{55} &= r_{86} =e^{h_{55}u}\,,\, r_{73} = r_{37} =e^{h_{33}u}\,,\, r_{68} = e^{h_{88}u}\,,\\
r_{13} &= \tfrac{h_{13}}{h_{44}-h_{55}}e_{13}\left(u\right),\,
r_{79}=\tfrac{h_{39}}{h_{44}-h_{55}}e_{13}\left(u\right)\\
r_{17} &= \tfrac{h_{17}}{h_{44}-h_{55}}e_{71}\left(u\right),\,
r_{39} = \tfrac{h_{79}}{h_{44}-h_{55}}e_{71}\left(u\right)\\
r_{26} &= \tfrac{h_{46}}{h_{44}-h_{55}}e_{45}\left(u\right),
\end{split}
\end{equation}
and finally
\begin{equation}
\label{eq:uppertr_rmatrix2}
\begin{split}
r_{19} &= \frac{e^{-h_{33}u}}{2\left(h_{44}-h_{55}\right)^2}\left[ \left(h_{13}h_{39} +h_{17}h_{79}\right) e_{31}^2\right. \\&\quad -\left.h_{19}(h_{44}-h_{55})\left(e^{2h_{33}u}-e^{2h_{11}u}\right)  \right]\,.
\end{split}
\end{equation}
where we have defined the combinations
\begin{equation}
\label{eq:uppertr_rmatdef}
\begin{split}
e_{ij}\left(u\right)&=e^{h_{ii}u}-e^{h_{jj}u},
\end{split}
\end{equation}
for notational convenience.

\section{More integrable Hamiltonians}
In this appendix, we present additional findings on novel integrable Hamiltonians mentioned in the main sections. For further examples, please refer to the attached Mathematica file.
\subsection{Hamiltonians arising from upper-triangular ansatz}\label{app:uppertr_hamil_19v} 
We now show two families, denoted $U_1,\,U_2$, of integrable Hamiltonians with 19 non-zero entries, obtained on starting from the upper-triangular ansatz in equation~\eqref{eq:uppertr_ansatz},
\begin{equation}
   H_{U_1/U_2} = \begin{pmatrix}
* & 0 & * & 0 & 0 & 0 & * & 0 & 0 \\
0 & * & 0 & * & 0 & 0 & 0 & * & 0 \\
0 & 0 & * & 0 & 0 & 0 & * & 0 & 0 \\
0 & * & 0 & * & 0 & * & 0 & 0 & 0 \\
0 & 0 & 0 & 0 & * & 0 & 0 & 0 & 0 \\
0 & 0 & 0 & 0 & 0 & * & 0 & * & 0 \\
0 & 0 & * & 0 & 0 & 0 & * & 0 & 0 \\
0 & 0 & 0 & 0 & 0 & * & 0 & * & 0 \\
0 & 0 & 0 & 0 & 0 & 0 & 0 & 0 & *
\end{pmatrix}
\end{equation}
Both families have all but one generically unequal entries. The equal non-zero entries in both families satisfy  the following linear relations
\begin{equation}
\begin{split}
h_{17} &=h_{28}= -h_{46} =-h_{13},\quad h_{42} = h_{68} = h_{37}, \\
    h_{33} &= h_{11} - h_{37},\quad
    h_{66} = -h_{37} - h_{88}, \\
     h_{22} &= 2h_{11} - h_{37} + h_{88}, \quad h_{44} = -2h_{11} - h_{88}, \\
    h_{77} &=-h_{99}= -h_{11},\quad
    h_{24} = h_{86} = h_{73}=2h_{11}\,,
\end{split}
\end{equation}
while the two classes $U_1,U_2$ are differentiated by their $h_{55}$ entry:
\begin{equation}
    U_1: \quad h_{55} = h_{11}\,,\qquad U_2: \quad h_{55}=-h_{11}-h_{37}\,.
\end{equation}

To illustrate the non-triviality of the varieties obtained, here is another example with 19 non-zero Hamiltonian entries
\begin{equation}
    H = \begin{pmatrix}
* & 0 & 0 & 0 & 0 & 0 & * & * & 0 \\
0 & * & 0 & 0 & 0 & 0 & * & * & * \\
0 & 0 & * & 0 & 0 & 0 & 0 & 0 & 0 \\
0 & 0 & 0 & * & 0 & 0 & * & * & 0 \\
0 & 0 & 0 & 0 & * & 0 & * & * & 0 \\
0 & 0 & 0 & 0 & 0 & * & 0 & 0 & * \\
0 & 0 & 0 & 0 & 0 & 0 & * & 0 & 0 \\
0 & 0 & 0 & 0 & 0 & 0 & 0 & * & 0 \\
0 & 0 & 0 & 0 & 0 & 0 & 0 & 0 & *
\end{pmatrix}
\end{equation}
In the gauge \(h_{58} = h_{69} =1\) and independent parameters \(\alpha,\beta,\gamma,\rho,\sigma,\tau\)
defined as
\begin{equation}
\label{eq:bigvariety_a}
\begin{split}
h_{11}=\alpha,\quad h_{28}=\rho,&\quad h_{29} =\sigma,\quad h_{59}=\tau\,,\\
h_{77} = \beta +\alpha\,,&\quad  h_{47} = 1+\gamma\,.
\end{split}
\end{equation}
The remaining non-vanishing elements of the Hamiltonian are parametrized as
\begin{equation}
\label{eq:bigvariety_b}
\begin{split}
h_{22}= h_{44}=h_{55}=h_{99}=\alpha \,,\\ h_{33}= h_{66} = \alpha-\beta\,,\\
h_{26} = \frac{\sigma\gamma \beta}{\gamma -\tau\beta}\,,
h_{27} = \frac{\rho\left(\gamma-\tau\beta\right)}{\sigma\beta}\,,\\
h_{17} = -\frac{\gamma\left(2\rho\left(\gamma-\tau\beta\right)+\sigma\beta\left(2\gamma-\tau\beta\right)\right)}{\tau\beta\left(\gamma+\tau\beta\right)}\,.
\end{split}
\end{equation}

\subsection{19-vertex ice-rule Hamiltonian}\label{app:ice-rule-example2}

R-matrices satisfying the ice-rule ansatz follow the constraint

\begin{equation}\label{eq:ice-rule}
R_{\mu\nu}^{\alpha\beta}=0,\quad \textrm{if} \quad \alpha+\beta\neq \mu+\nu\,,
\end{equation}

where the indices $\alpha,\beta,\mu,\nu$ all range in the triplet of values $\{-1,0,1\}$, corresponding to the local 3 dimensional representations of the ``in"  and ``out" states. Looking for solutions in this restricted setting, we obtained several novel solutions, one of which is discussed in equation~\eqref{ice-rule-example}. Another 3-parameter family of integrable Hamiltonians, parametrized by ($h_{35},h_{37},h_{44}$) is presented below
\begin{equation}
\begin{split}
h_{11} &= h_{99} = 1\,, \quad 
h_{88} = h_{44}\,, \\
h_{22} &= h_{66}= 2 + \frac{h_{35}^2}{h_{37}} - 2h_{37} - h_{44}\,,\\
h_{24} &= h_{42}= h_{68} = h_{86}= -h_{73} = -h_{37}\,, \\
h_{33} &= 3 + \frac{2h_{35}^2}{h_{37}} - 3h_{37} - 2h_{44}\,, \\
h_{53} &= h_{57} = h_{75} = h_{35}\,,\quad
h_{55} = 1 + \frac{h_{35}^2}{h_{37}} - 2h_{37}\,, \\
h_{77} &= -1 + h_{37} + 2h_{44}\,.
\end{split}
\end{equation}

\bibliography{apssamp}

\end{document}